%% file: dyb-minos-sterile.tex
\documentclass[aps,prl,showpacs,showkeys,amsmath,amssymb,
twocolumn,
floatfix,
preprintnumbers,
superscriptaddress
]{revtex4-1}
\usepackage{graphicx}
\usepackage{siunitx}
\usepackage{times}
\usepackage{verbatim}
\usepackage{multirow}

\usepackage[normalem]{ulem}
\usepackage{accents}
\usepackage[tight]{units} 

\newcommand {\delm}[1]{\ensuremath{\Delta m^{2}_{\rm{#1}}}}
\newcommand {\cls}{\ensuremath{\rm{CL_{\it s}}}}
\newcommand {\clb}{\ensuremath{\rm{CL_{\it b}}}}
\newcommand {\clsb}{\ensuremath{\rm{CL_{\it s+b}}}}

\begin{document}

\title{Limits on Active to Sterile Neutrino Oscillations from Disappearance Searches in the MINOS, Daya~Bay, and Bugey-3 Experiments}

\input{merged_minos_dyb_authors.tex}
\preprint{FERMILAB-PUB-16-234-ND}

\date{\today}

\begin{abstract}
Searches for a light sterile neutrino have been performed independently by 
the MINOS and the Daya Bay experiments using the muon (anti)neutrino and 
electron antineutrino disappearance channels, respectively. 
In this Letter, results from both experiments are combined with those from the Bugey-3 reactor neutrino experiment to constrain oscillations into light sterile neutrinos. The three experiments are sensitive to complementary regions of parameter space, enabling the combined analysis to probe regions allowed by the Liquid Scintillator Neutrino Detector (LSND) and MiniBooNE experiments in a minimally extended four-neutrino flavor framework. Stringent limits on $\sin^2 2\theta_{\mu e}$ are set over 6 orders of magnitude in the sterile mass-squared splitting $\Delta m^2_{41}$.  The sterile-neutrino mixing phase space allowed by the LSND and MiniBooNE experiments is excluded for $\Delta m^2_{41} < 0.8$ eV$^2$ at 95\% CL$_s$.
\end{abstract}

\maketitle

The discovery of neutrino flavor oscillations~\cite{Fukuda:1998mi,Ahmad:2001an} marked a crucial milestone in the history of particle physics. It indicates neutrinos undergo mixing between flavor and mass eigenstates and hence carry nonzero mass. It also represents the first  evidence of physics beyond the standard model of particle physics.
Since then, neutrino oscillations have been confirmed and precisely measured with data from 
natural (atmospheric and solar) and man-made (reactor and accelerator) 
neutrino sources.

The majority of neutrino oscillation data available can be well
described by a three-flavor neutrino 
model~\cite{Pontecorvo:1957cp,Pontecorvo:1967fh,Maki:1962mu} in agreement 
with precision electroweak measurements from 
collider experiments~\cite{PDG2014,ALEPH:2005ab}. A few experimental results, however, 
including those from the Liquid Scintillator Neutrino Detector (LSND)~\cite{Aguilar:2001ty} and MiniBooNE~\cite{Aguilar-Arevalo:2013pmq} experiments, cannot be 
explained by three-neutrino mixing. 
Both experiments observed an electron antineutrino excess in a muon antineutrino beam
over short baselines, suggesting mixing with a new neutrino 
state with mass-squared splitting $\Delta m^2_{41}\gg|\Delta m^2_{32}|$, 
where $\Delta m^2_{ji}\equiv m^2_{j} - m^2_i$, and $m_i$  is the mass of the $i\mathrm{th}$ mass eigenstate. Precision electroweak measurements exclude standard couplings of 
this additional neutrino state for masses up to 
half the $Z$-boson mass, so that states beyond the known three active 
states are referred to as {\it sterile}. New light neutrino states would open a new sector in particle physics;  thus, confirming or refuting these results is at the forefront of neutrino physics research.

Mixing between one or more light sterile neutrinos and the active 
neutrino flavors would have discernible effects on neutrino 
oscillation measurements. Oscillations from muon to electron (anti)neutrinos driven by a sterile neutrino require electron and muon neutrino flavors to couple to the additional neutrino 
mass eigenstates. Consequently, 
oscillations between active and sterile states will also necessarily result in the disappearance of muon (anti)neutrinos, as well as 
of electron (anti)neutrinos~\cite{Okada:1996kw,Bilenky:1996rw}, independently of the sterile neutrino model considered~\cite{Giunti:2015mwa, Kopp:2013vaa}. 

In this Letter, we report results from a joint analysis developed in parallel to the independent sterile neutrino searches from the Daya Bay~\cite{dyb_6+8_sterile} and the MINOS~\cite{newminossterile} experiments. In this analysis, the measurement of muon (anti)neutrino 
disappearance by the MINOS experiment is combined with electron antineutrino disappearance measurements from the Daya Bay and Bugey-3~\cite{bugey-3:paper} experiments using the signal confidence level $\left(\cls \right)$~method~\cite{Read:2002hq,Junk1999435}.  The combined results are analyzed in light of the muon (anti)neutrino to electron (anti)neutrino appearance 
indications from the LSND~\cite{Aguilar:2001ty} and MiniBooNE~\cite{Aguilar-Arevalo:2013pmq} experiments. 
The independent MINOS, Daya Bay, and Bugey-3 results are all 
obtained from disappearance measurements and therefore are insensitive to 
{\it CP}-violating effects due to mixing between the three active flavors. Under the assumption of {\it CPT} invariance, the combined results shown  constrain both neutrino and antineutrino appearance. 

The results reported here required several novel improvements developed independently from the Daya Bay-only~\cite{dyb_6+8_sterile} and MINOS-only~\cite{newminossterile} analyses, specifically: a full reanalysis of the MINOS data to search for sterile neutrino mixing, based on the \cls~method, a  \cls-based analysis of the Bugey-3 results taking into account new reactor flux calculations and the Daya Bay experiment's reactor flux measurement, the combination of the Daya Bay results with the Bugey-3  results taking into account correlated systematics between the experiments, and, finally, the combination of the Daya Bay + Bugey-3 and MINOS results to place stringent constraints on electron neutrino and antineutrino appearance driven by sterile neutrino oscillations. 

We adopt a minimal extension of the three-flavor neutrino 
model by including one sterile flavor and one additional mass 
eigenstate. This 3+1 sterile neutrino scenario is referred to as the {\it four-flavor model} in the text.  In this model, the muon to electron neutrino appearance probability $P_{\nu_\mu\rightarrow\nu_e}(L/E)$ as a function of the propagation length $L$, divided by the neutrino energy $E$, can be expressed using a 4$\times$4 unitary mixing matrix, $U$, by

\begin{equation}\label{eq:PueFullnoCP}
P_{\nu_\mu\rightarrow\nu_e}(L/E) =  \left |\sum_{i} U_{li}U^{*}_{l'i}e^{-i(m_{i}^2/2E)L} \right | ^2 
\end{equation}

In the region where $\Delta m^2_{41}$ $\gg$ $|\Delta m^2_{32}|$ and for short baselines $\left(\left(\Delta m^2_{32}L / 4E\right) \sim 0\right)$, 
Eq.~\eqref{eq:PueFullnoCP} can be simplified to   

\begin{equation}\label{eq:PueApproxMtrx}
P_{\nu_\mu\rightarrow\nu_e}(L/E) \approx
4 |U_{e 4}|^2|U_{\mu 4}|^2\sin^2\left(\frac{\Delta m^2_{41}L}{4E}\right) \approx P_{\bar{\nu}_\mu\rightarrow \bar{\nu}_e}.
\end{equation}

A nonzero amplitude for the appearance probability, $4|U_{e 4}|^2|U_{\mu 4}|^2$, is a possible explanation for the MiniBooNE and LSND results. The matrix element
$|U_{e4}|^2$ can be constrained with measurements of electron antineutrino 
disappearance, as in the Daya 
Bay~\cite{dyb_6+8_sterile} and Bugey-3~\cite{bugey-3:paper} 
experiments. Likewise, $|U_{\mu4}|^2$ can be constrained with measurements of 
muon neutrino and antineutrino disappearance, as in the 
MINOS~\cite{newminossterile} experiment. 
For these experiments, the general four-neutrino survival probabilities $P_{\bar\nu_e\rightarrow\bar\nu_e}(L/E)$ and $P_{\overset{(-)}\nu\!\!_\mu\rightarrow\overset{(-)}\nu\!\!_\mu}(L/E)$ are 

\begin{align}
P_{\overline\nu_e\rightarrow\overline\nu_e}(L/E) = & \,1-4\sum_{k>j}|U_{ek}|^2|U_{ej}|^2\sin^2\left(\frac{\Delta m^2_{kj}L}{4E}\right), \label{eq:PeeFullU}\\ 
P_{\overset{(-)}\nu\!\!_\mu\rightarrow\overset{(-)}\nu\!\!_\mu}(L/E) = & \,1-4\sum_{k>j}|U_{\mu k}|^2|U_{\mu j}|^2\sin^2\left(\frac{\Delta m^2_{kj}L}{4E}\right). \label{eq:PuuFullU}
\end{align}

The mixing matrix augmented with one sterile state can be parametrized by 

$U=R_{34}R_{24}R_{14}R_{23}R_{13}R_{12}$~\cite{Harari:1986xf}, where $R_{ij}$ is the rotational matrix for the mixing angle $\theta_{ij}$, yielding  
\begin{eqnarray}
\begin{split}
|U_{e4}|^2 =& \,\,\sin^2\theta_{14}, \\
|U_{\mu4}|^2 =& \,\,\sin^2\theta_{24}\cos^2\theta_{14},\\
4|U_{e4}|^2|U_{\mu4}|^2 =&\,\,\sin^22\theta_{14}\sin^2\theta_{24} \equiv \sin^22\theta_{\mu e}. \label{eq:DisapToApp}
\end{split}
\end{eqnarray}

Searches for sterile neutrinos are carried out by using the reconstructed energy spectra to look for evidence of oscillations driven by the sterile mass-squared difference $\delm{41}$. For small values of $\delm{41}$, corresponding to slow oscillations, the energy-dependent shape of the oscillation probability could be measured in the reconstructed energy spectra. For large values corresponding to rapid oscillations, an overall reduction in neutrino flux would be seen.

The  \cls~method~\cite{Read:2002hq,Junk1999435} is a two-hypothesis test 
that compares the three-flavor (null) hypothesis (labeled 3$\nu$) to an 
alternate four-flavor hypothesis (labeled 4$\nu$). To determine if the four-flavor 
hypothesis can be excluded, we construct the test statistic 
$\Delta \chi^2 = \chi^2_{4\nu} - \chi^2_{3\nu}$, where $\chi^2_{4\nu}$ is the $\chi^2$ 
value resulting from a fit to a four-flavor hypothesis, and 
$\chi^2_{3\nu}$ is the $\chi^2$ value from a fit to the three-flavor 
hypothesis. The $\Delta \chi^2$ observed with data, $\Delta \chi^2_\mathrm{obs}$, is compared 
to the $\Delta \chi^2$ distributions expected if the three-flavor hypothesis is true, or
the four-flavor hypothesis is true. To quantify this, we construct
  \begin{eqnarray}
    \begin{split}
      \clb =&\,\, P( \Delta \chi^2 \geq \Delta \chi^2_\mathrm{{obs}} | 3\nu ), \\
      \clsb =&\,\, P( \Delta \chi^2 \geq \Delta \chi^2_\mathrm{{obs}} | 4\nu ), \\
      \cls =&\,\, \frac{\clsb}{\clb}, \label{eq:clsDef}
    \end{split}
  \end{eqnarray}
over a grid of $(\sin^2 2\theta_{14}, \Delta m^2_{41})$ points for the Daya Bay + Bugey-3 experiments and a grid of $(\sin^2 \theta_{24}, \Delta m^2_{41})$ for the MINOS experiment. 
\clb\ measures consistency with the three-flavor hypothesis, 
and \clsb\ measures the agreement with the four-flavor hypothesis.
The alternate hypothesis is excluded at the $\alpha$ confidence level if 
$\cls \leq 1 - \alpha$. The construction of \cls\ ensures that even if \clsb\ 
is small, indicating disagreement with the four-flavor hypothesis, 
this hypothesis can only be excluded when \clb\ is large, indicating
consistency with the three-flavor hypothesis. Thus, the \cls\ construction
ensures the four-flavor hypothesis can only be excluded if the experiment is sensitive to it.

Calculating \clb\ and \clsb\ can be done in two ways. The first method is the Gaussian 
\cls\ method~\cite{CLs:Qian2014}, which uses two Gaussian $\Delta \chi^2$ distributions. 
The first distribution is obtained by fitting toy Monte Carlo (MC) data assuming the 
three-flavor hypothesis is true, thus labeled as $\Delta \chi^2_{3\nu}$.
The second distribution is obtained by assuming the four-flavor hypothesis is
true ($\Delta \chi^2_{4\nu}$).
The mean of each distribution is obtained from a fit to the Asimov data set, an infinite statistics sample where the relevant parameters are set to best-fit values for each hypothesis~\cite{asimov}. The Gaussian width for the Asimov data set is derived analytically. In the second method, 
the distributions of $\Delta \chi^2$ are approximated by MC simulations of pseudoexperiments. 
The Gaussian method is used to obtain the Daya Bay and Bugey-3 combined results, while the second method is used to obtain the MINOS results.

The MINOS experiment~\cite{Michael:2008bc} operates two functionally equivalent detectors separated by $734\,$km. The detectors sample the NuMI neutrino beam~\cite{Adamson:2015dkw}, which yields events with an energy spectrum that peaks at about 3\,GeV. 
Both detectors are magnetized steel and scintillator calorimeters, with the 1\,kton Near Detector (ND) situated 1\,km downstream of the NuMI production target, and the 5.4\,kton Far Detector (FD) located at the Soudan Underground Laboratory~\cite{Michael:2008bc}. The analysis reported here uses data from an exposure of $10.56\times\nolinebreak10^{20}\,$protons on target, for which the neutrino beam composition is 91.8\%~$\nu_\mu$, 6.9\% $\bar{\nu}_\mu$, and 1.3\% ($\nu_e+\bar{\nu}_e$).

To look for sterile neutrino mixing, the MINOS experiment uses the reconstructed energy spectra in the ND and FD of both charged-current (CC) and neutral-current (NC) neutrino interactions. The sterile mixing signature differs depending on the range of \delm{41} values considered. For $\Delta m^2_{41} \in (0.005,0.05)\ {\rm eV}^2$, the muon neutrino CC spectrum in the FD would display deviations from three-flavor oscillations. For rapid oscillations driven by $\Delta m^2_{41} \in (0.05,0.5)\ {\rm eV}^2$, the combination of finite detector energy resolution and rapid oscillations at the FD location would result in an apparent event rate depletion between the ND and FD. For larger sterile neutrino masses, corresponding to \delm{41}$>$\unit[0.5]{eV$^{2}$}, oscillations into sterile neutrinos would distort the ND CC energy spectrum. Additional sensitivity is obtained by analyzing the reconstructed energy spectrum for NC candidates. The NC cross sections and interaction topologies are identical for all three active neutrino flavors, rendering the NC spectrum insensitive to standard oscillations, but mixing with a sterile neutrino state would deplete the NC energy spectrum at the FD, as the sterile neutrino would not interact in the detector. For large sterile neutrino masses, such depletion would also be measurable at the ND.

The simulated FD-to-ND ratios of the reconstructed energy spectra for $\nu_\mu$ CC and NC selected events, including four-flavor oscillations for both the ND and FD, are fit to the equivalent FD-to-ND ratios obtained from data~\cite{newminossterile}. Current and previous results of the MINOS sterile neutrino searches, along with further analysis details, are described in Refs.~\cite{newminossterile,Adamson:2011ku,Adamson:2010wi,Adamson:2008jh}. The  MINOS experiment employs the Feldman-Cousins ordering principle~\cite{Feldman:1997qc} in obtaining exclusion limits in the four-flavor parameter space. However, this approach requires a computationally impractical joint fit to be consistent, since it requires minimizing $\chi^{2}$ over $\Delta m^{2}_{41}$, a shared parameter between the MINOS and Daya Bay + Bugey-3 experiments. Thus, the \cls\ method described above is used. 

While the MINOS experiment does not have any sensitivity to $\sin^2 \theta_{14}$, there is a small
sensitivity to $\sin^2 \theta_{34}$ due to the inclusion of the NC channel.
During the fit, $\sin^2 \theta_{34}$ is allowed to vary freely in addition to $\Delta m^2_{32}$ and $\sin^2 \theta_{23}$, while $\sin^2 \theta_{24}$ and $\Delta m^2_{41}$ are held fixed to define the particular four-flavor hypothesis that is being tested. Since the constraint on $\sin^2\theta_{34}$ is relatively weak, the distribution of $\Delta \chi^2$ deviates from the normal distribution and 
the Gaussian \cls\ method cannot be used. The $\Delta \chi^2_{3\nu}$ and $\Delta \chi^2_{4\nu}$ distributions are constructed by fitting pseudoexperiments. 
 
\begin{figure}[htbp]                                                                                                                                        
\includegraphics[width=\columnwidth]{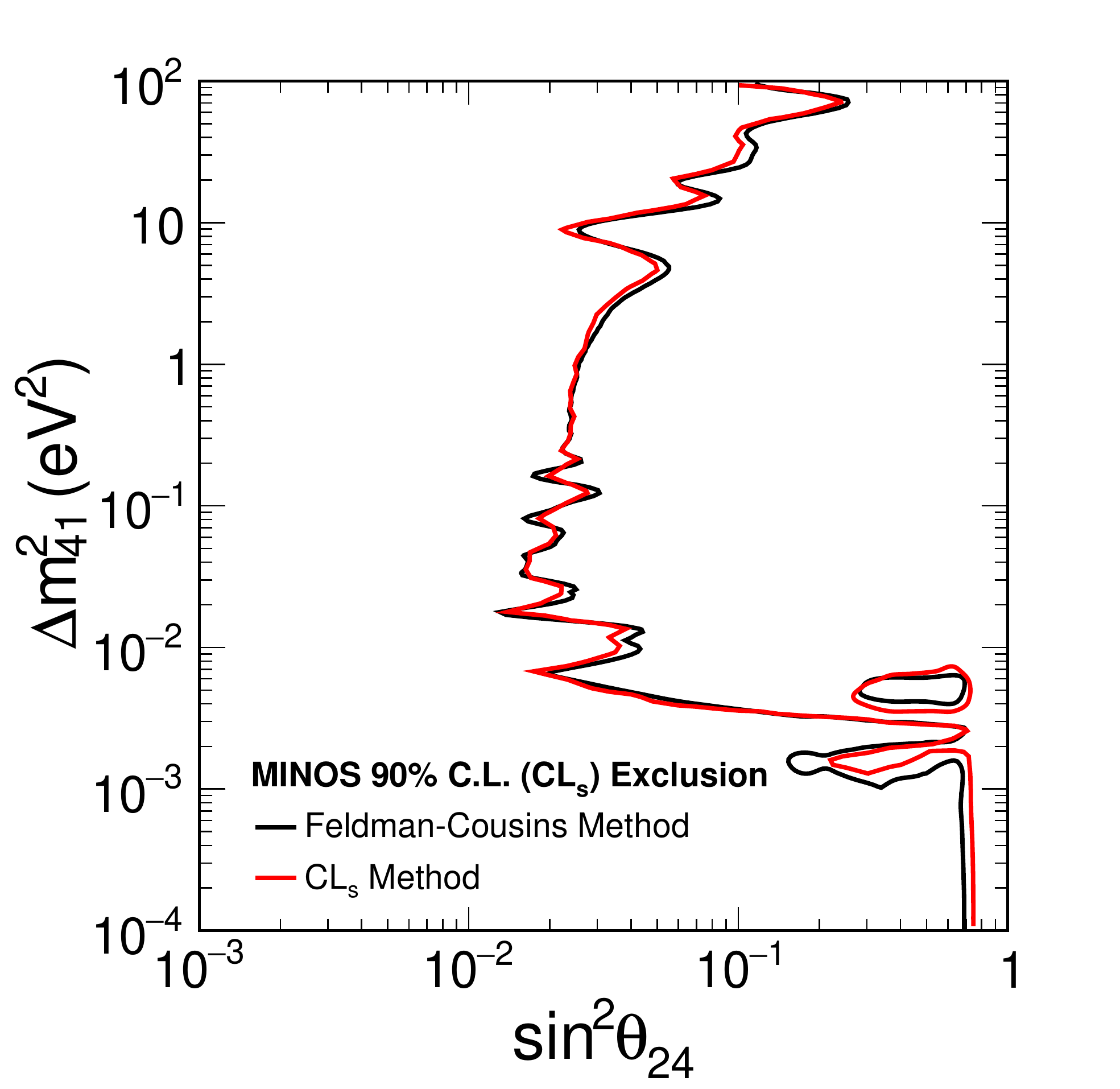}                            
\caption{\label{fig:minoscls} Comparison of the MINOS 90\% CL contour using the Feldman-Cousins method~\cite{newminossterile} and the \cls\ method. The region to the right of the curve is excluded at the 90\% CL (\cls).}
\end{figure}

In the three-flavor case, pseudoexperiments are simulated using the same parameters listed in Ref.~\cite{newminossterile}, {\it i.e.} $\sin^2 \theta_{12}=0.307$, $\Delta m^2_{21}=7.54\times 10^{-5}$\,eV$^2$ based on a global fit to neutrino data~\cite{Fogli:2012ua}, 
and $\sin^2 \theta_{13}= 0.022$ based on a weighted average of results from reactor experiments~\cite{DYB:8AD, RENO:2015ksa, Abe:2014bwa}. 
For the atmospheric oscillation parameters, equal numbers of pseudoexperiments are simulated in the upper and lower octant ($\sin^2\theta_{23} = 0.61$ and $\sin^2\theta_{23} = 0.41$, respectively), with $|\Delta m^2_{32}|=2.37\times 10^{-3}$\,eV$^2$, based on the most recent MINOS results~\cite{Adamson:2014vgd}. The uncertainties on solar oscillation parameters have negligible effect on the analysis, so fixed values are used. 
In the four-flavor case, $|\Delta m^2_{32}|$, $\sin^2 \theta_{23}$, and $\sin^2 \theta_{34}$ are taken from fits to data at each $(\sin^2 \theta_{24},\Delta m^2_{41})$ grid point. In both the three- and four-flavor cases, half of the pseudoexperiments are generated in each mass hierarchy. A comparison of MINOS exclusion contours obtained using the Feldman-Cousins procedure~\cite{newminossterile} with those obtained using the \cls\ method is shown in Fig.~\ref{fig:minoscls}.   Note that if $\delm{41} = 2\delm{31}$ or $\delm{41} \ll \delm{31}$ and $\sin^2\theta_{23} = \sin^2\theta_{34} = 1$, $\theta_{24}$ can take on the role normally played by $\theta_{23}$.  In these cases, the four-flavor model is degenerate with the three-flavor model, leading to regions of parameter space that cannot be excluded.

The Daya Bay experiment measures electron antineutrinos via inverse $\beta$ decay (IBD): $\bar\nu_e+p\rightarrow e^+ + n$. The antineutrinos are  produced by six reactor cores and detected in eight identical Gd-doped liquid-scintillator antineutrino detectors (ADs)~\cite{DayaBay:detectors} in three underground experimental halls (EHs).
The flux-averaged baselines for EH1, EH2, and EH3 are 520, 570, and 1590\,m, respectively.
The target mass in each of the two near EHs is 40 tons, and that in the far EH is 80 tons.
Details of the IBD event selection, background estimates, and assessment of systematic uncertainties can be found in Refs.~\cite{An:2013zwz,DYB:8AD}. By searching for distortions in the $\bar\nu_e$ energy 
spectra, the experiment is sensitive to $\sin^22\theta_{14}$ for a mass-squared splitting $\Delta m^2_{41} \in (0.0003,0.2)\,{\rm eV}^2$.
For $\Delta m^2_{41}\textgreater\,0.2\,{\rm eV}^2$, spectral 
distortions cannot be resolved by the detector. 
Instead, the measured antineutrino flux can be compared 
with the predicted flux to constrain the sterile neutrino parameter space. 
Recently, the Daya Bay Collaboration published its measurement of the overall antineutrino flux~\cite{An:2015nua}.
The result is consistent with previous measurements at short baselines, which prefer 5\% lower values than the latest calculations~\cite{Huber:2011wv,Mueller:2011nm}, a deficit commonly referred to as the reactor antineutrino anomaly~\cite{anom}. However, the reactor spectrum measurement from the Daya Bay Collaboration~\cite{An:2015nua} (and from the RENO Collaboration~\cite{RENO:2015ksa} and the Double Chooz Collaboration~\cite{Abe:2014bwa}) shows clear discrepancies with the latest calculations, which indicates an underestimation of their uncertainties. The uncertainties on the antineutrino flux models for this analysis are increased to 5\% from the original 2\% as suggested by Refs.~\cite{Hayes:2013wra,Vogel:2016ted}.
The Daya Bay Collaboration has recently updated the sterile 
neutrino search result in Ref.~\cite{dyb_6+8_sterile} with limits on 
$\sin^22\theta_{14}$ improved by about a factor of two with respect 
to previous results~\cite{DYB:sterile6AD}. This data set is used in producing the combined results presented here.

Two independent sterile neutrino search analyses are conducted by Daya Bay. 
The first analysis uses the predicted $\bar\nu_e$ spectrum  to generate the 
predicted prompt spectrum for each antineutrino detector simultaneously, taking into account detector effects such as energy resolution, 
nonlinearity, detector efficiency, and oscillation parameters 
described in~\cite{DYB:8AD}.  A log-likelihood 
function is constructed with nuisance parameters to include the 
detector-related uncertainties and a covariance matrix to incorporate 
the uncertainties on reactor neutrino flux prediction. 
The Gaussian CL$_s$ method is used to calculate the excluded region. The second analysis uses the observed spectra 
at the near sites to predict the far site spectra to further reduce the dependency 
on reactor antineutrino flux models. 
Both analyses yield consistent results~\cite{dyb_6+8_sterile}.

\begin{figure}[htbp]
\includegraphics[width=\columnwidth]{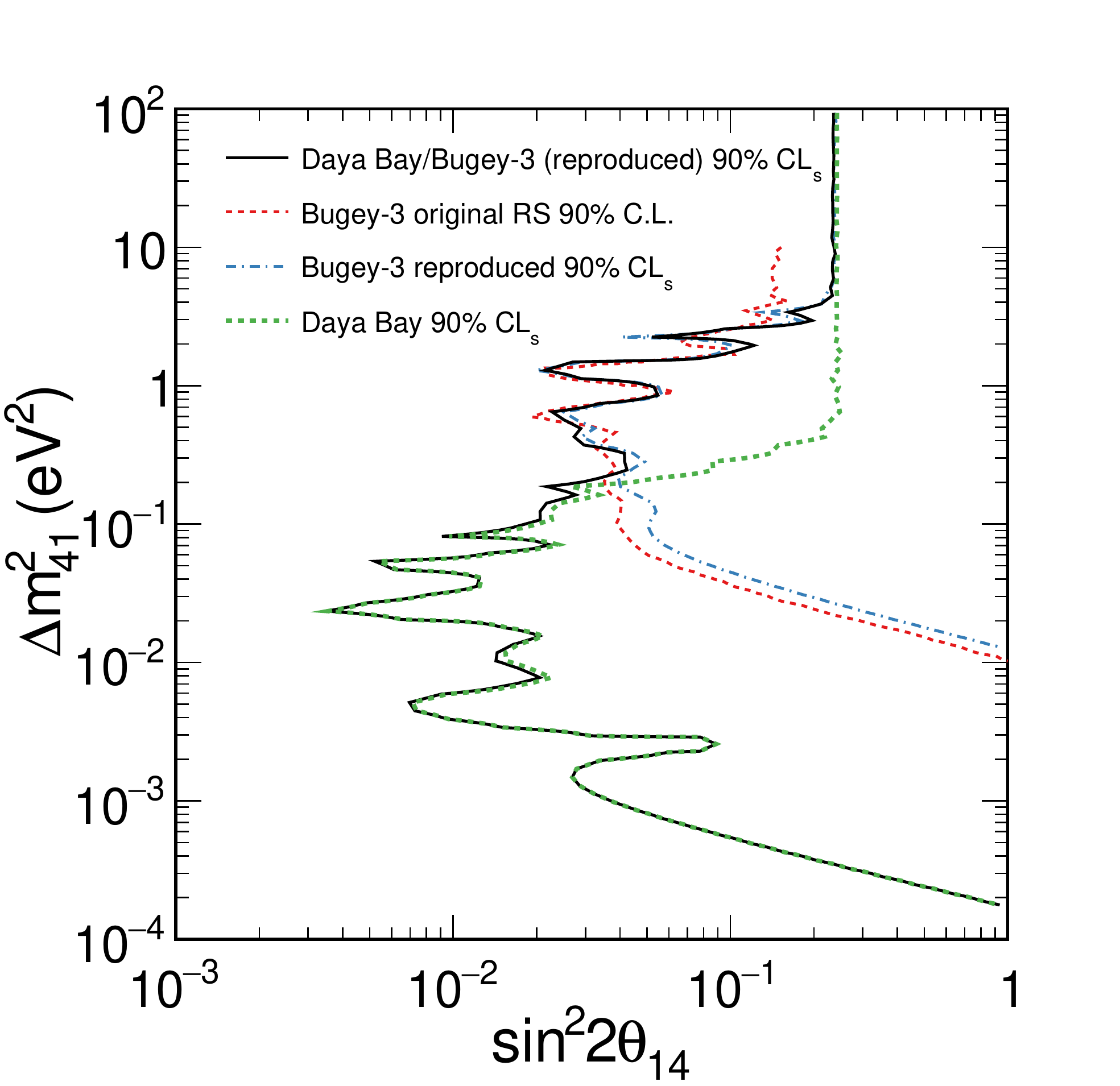}
\caption{\label{fig:DYBBGY} Excluded regions for the original Bugey-3 raster scan (RS) result~\cite{bugey-3:paper}, for the reproduced Bugey-3 with adjusted fluxes, for the Daya Bay result~\cite{dyb_6+8_sterile}, and for the combined Daya Bay and reproduced Bugey-3 results. The region to the right of the curve is excluded at the 90\% CL$_s$.}
\end{figure}

The Bugey-3 experiment was performed in the early 1990s and its main goal was to search for neutrino oscillations 
using reactor antineutrinos.  In this experiment, two $^{6}$Li-doped liquid scintillator detectors measured 
$\bar{\nu}_{e}$ generated from two reactors at three different baselines 
(15, 40 and 95\,m)~\cite{bugey-3:paper}. The Bugey-3 experiment detected IBD interactions with the recoil neutron 
capturing on $^{6}$Li (n + $^{6}$Li $\to$ $^{4}$He + $^{3}$H + 4.8 MeV). Probing shorter 
baselines than the Daya Bay exeriment, the Bugey-3 experiment is sensitive to regions of parameter space with larger $\Delta m^2_{41}$ values. 

The original Bugey-3 results obtained using the raster scan technique 
are first reproduced employing a $\chi^2$ definition used in the original Bugey-3 
analysis~\cite{bugey-3:paper}:
\begin{equation}
\begin{split}
\chi^2 =
&\sum_i^3\sum_j^{N_i}\frac{\left\{\left[Aa_i+b(E_j-1.0)\right]R^\mathrm{pre}_{i,j}-R^\mathrm{obs}_{i,j}\right\}^2}{\sigma_{i,j}^2}\\
&+\sum_i^3\frac{(a_i-1)^2}{\sigma_{a_i}^2}+\frac{(A-1)^2}{\sigma_{A}^2}+\frac{b^2}{\sigma_{b}^2},
\end{split}
\end{equation}
where $A$ is the overall normalization, $a_i$ is the relative detection efficiency, $b$ is an empirical factor 
to include the uncertainties of the energy scale, $i$ represents the data from three baselines, and $j$ sums over the $N_i$ bins at each baseline.
The values of $\sigma_{a_i}$ and $\sigma_{b}$ are set at $0.014$\,MeV$^{-1}$ and $0.020$\,MeV$^{-1}$, respectively, according to the reported values in Ref.~\cite{bugey-3:paper}. 
The $\sigma_{i,j}$ are the statistical uncertainties.
The uncertainty on the overall normalization $\sigma_A$ is set to 5\% 
to be consistent with the constraint employed in the Daya Bay analysis~\cite{dyb_6+8_sterile}.
The ratio of the observed spectrum to the predicted unoscillated spectrum is denoted by $R^\mathrm{obs}_{i,j}$,
 while $R^\mathrm{pre}_{i,j}$ is the predicted ratio of the spectrum including oscillations to the one without oscillations.
To predict the energy spectra, the average fission fractions are used~\cite{bugey-4}, and the energy resolution is set to 5\% at 4.2\,MeV~\cite{bugey-3:paper} with a functional form similar to the Daya Bay experiment's. 
The predicted energy spectra are validated against the published Bugey-3 spectra~\cite{bugey-3:paper}.

In the Bugey-3 experiment, the change in the oscillation probability over the baselines of the 
detectors and the reactors is studied with MC simulations assuming 
that antineutrinos are uniformly generated in the reactor cores and uniformly measured in 
the detectors, and approximated by treating the baselines as normal distributions.
To achieve the combination with the Daya Bay experiment, two changes are made in the reproduced Bugey-3 analysis: 
the change in the cross section of the IBD process due to the updated neutron decay time~\cite{PDG2014} is applied, and  the antineutrino flux is adjusted from the ILL+Vogel model~\cite{ILL,vogel}  to that of Huber~\cite{Huber:2011wv} and Mueller~\cite{Mueller:2011nm}, for consistency with the prediction used by the Daya Bay experiment. These adjustments change the reproduced contour with respect to the original Bugey-3 one, in particular by reducing the sensitivity to regions with $\Delta m^2_{41}>3$\,eV$^2$, with less noticeable effects for smaller $\Delta m^2_{41}$ values.
The reproduced Bugey-3 limit on the sterile neutrino mixing, 
and the limit obtained by combining the Bugey-3 with the Daya Bay results
through a $\chi^2$ fit, with 
common overall normalization and oscillation parameters, are shown in Fig.~\ref{fig:DYBBGY}.

\begin{figure}[htbp]
\includegraphics[width=\columnwidth]{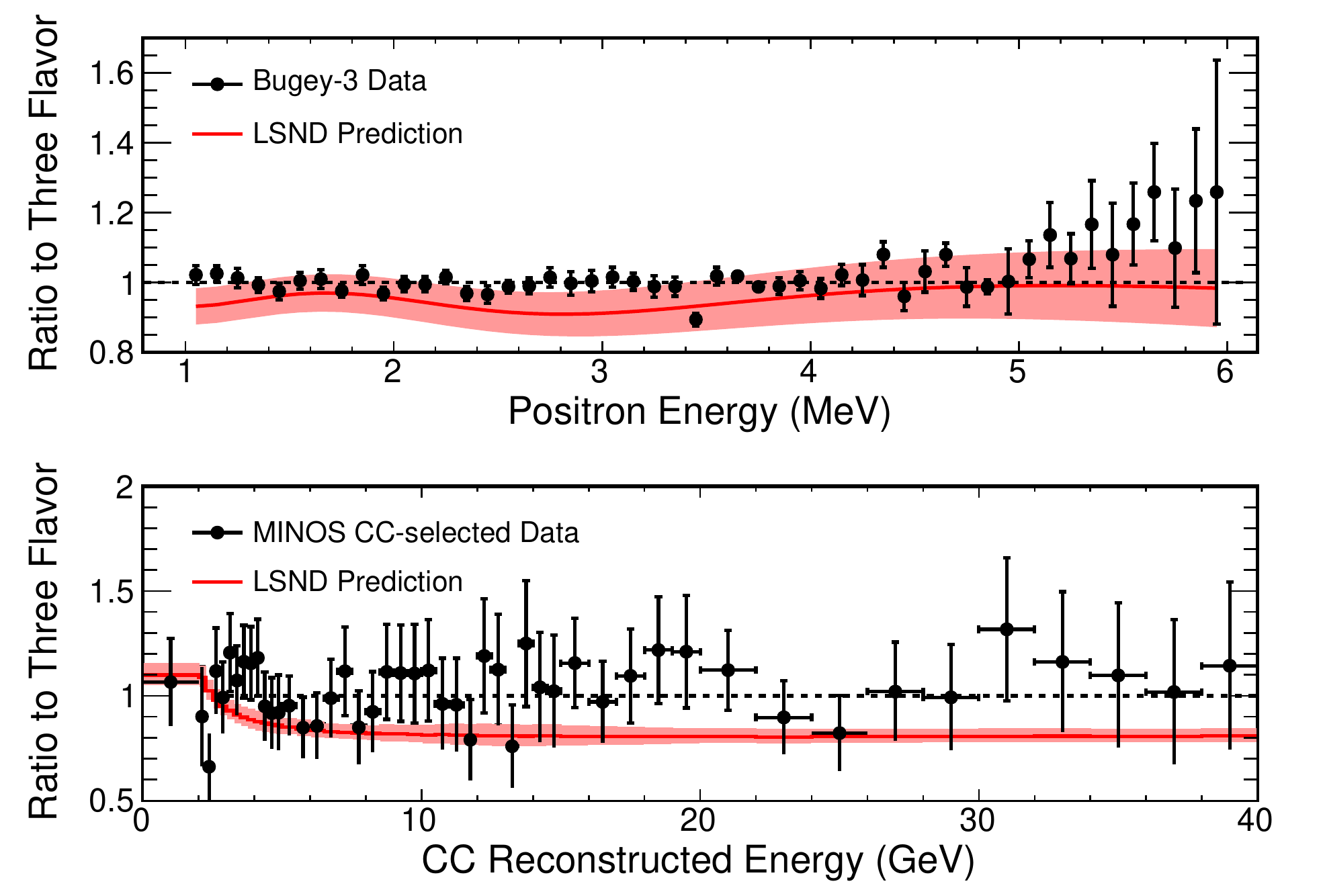}
\caption{\label{fig:LSNDComp} The top panel shows the ratio of the Bugey-3 15 m IBD data to a three-neutrino prediction, while the bottom panel shows the ratio of the MINOS FD-to-ND ratio data for CC events to a three-neutrino prediction. The red lines represent the four-flavor predictions at ($\Delta m^2_{41}$~$=$~1.2~eV$^2$, $\sin^2 2\theta_{\mu e}$~$=$~0.003). The shaded band displays the sizes of the systematic uncertainties. A value of $\sin^22\theta_{14} = 0.11$ is used for the Bugey-3 prediction so that when multiplied by the MINOS 90\% CL$_s$ limit on $\sin^2\theta_{24}$, it matches $\sin^2{2\theta_{\mu e}}=0.003$. A $\Delta\chi^2$ value of 48.2 is found between the data and this $4\nu$ prediction. Similarly, a value of $\sin^2\theta_{24} = 0.12$ is combined with the Bugey-3 90\% CL$_s$ limit on $\theta_{14}$ to produce the MINOS four-flavor prediction, resulting in $\Delta\chi^2 = 38.0$ between the data and the prediction.}
\end{figure}

Individually, the MINOS and Bugey-3 experiments are both sensitive to regions of parameter space allowed by the LSND measurement through constraints on $\theta_{24}$ and $\theta_{14}$, shown in Figs.~\ref{fig:minoscls} and~\ref{fig:DYBBGY}, respectively. We illustrate this sensitivity in Fig.~\ref{fig:LSNDComp}, which displays a comparison of the energy spectra for Bugey-3 and MINOS data to four-flavor ($4\nu$) predictions produced at the LSND best-fit point~\cite{Aguilar:2001ty} as an example. For Bugey-3, a $\Delta\chi^2$ value of 48.2 is found between the data and the four-flavor prediction. Taking equal priors between these two models, the posterior likelihood for $3\nu$ vs $4\nu$ is 1 vs $3.4\times10^{-11}$ in the Bayesian framework. For the MINOS experiment, a $\Delta\chi^2$ value of 38.0 is obtained between the data and the prediction. The posterior likelihood for $3\nu$ vs $4\nu$ is 1 vs $5.6\times10^{-9}$.

\begin{figure}[htbp]
\includegraphics[width=\columnwidth]{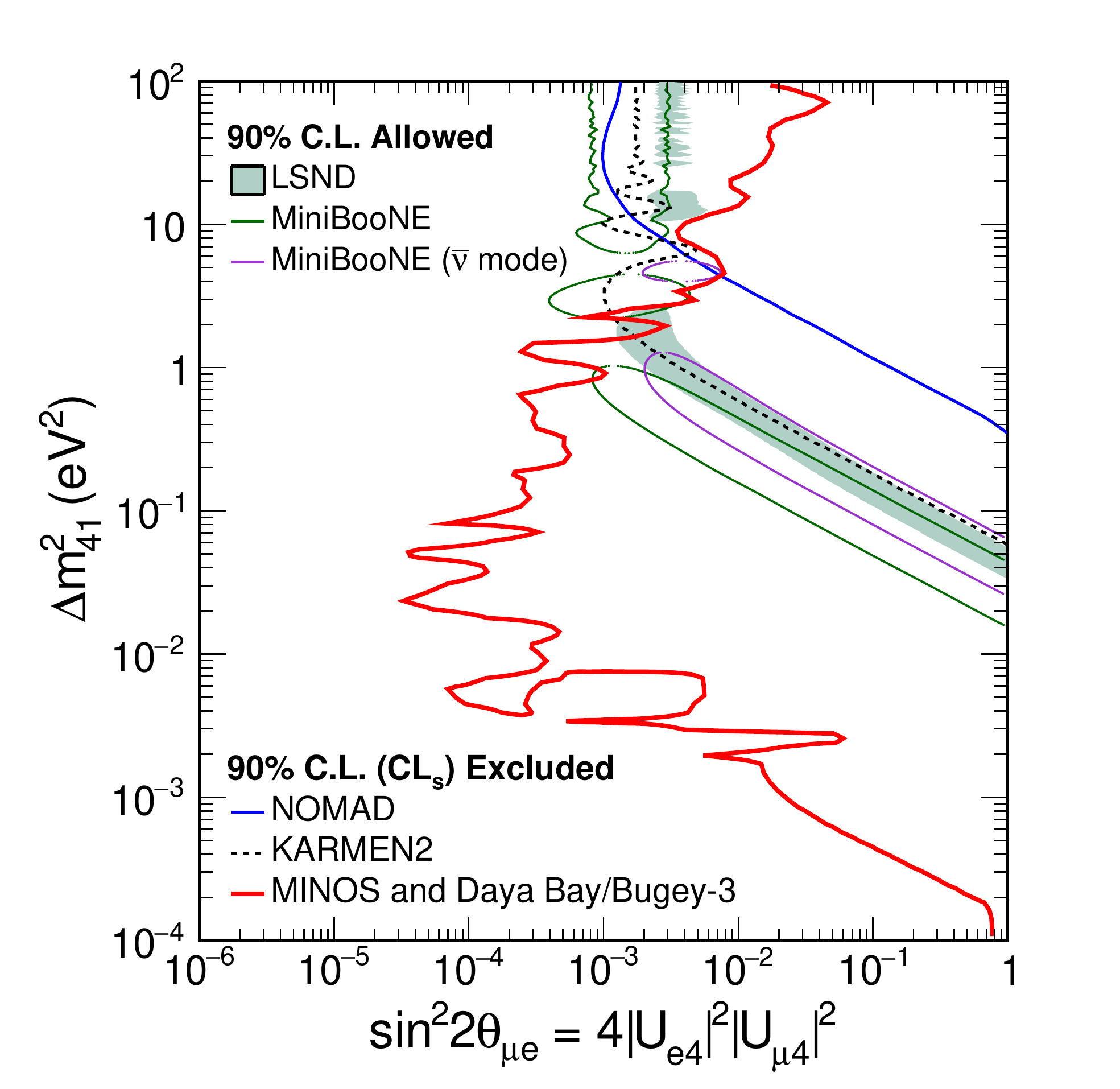}
\caption{\label{fig:combocls}MINOS and Daya Bay + Bugey-3 combined 90\% \cls\ limit on ${\sin}^22{\theta}_{\mu e}$ compared to the LSND and MiniBooNE 90\% CL allowed regions. 
Regions of parameter space to the right of the red contour are excluded. 
The regions excluded at 90\% CL by the KARMEN2 Collaboration~\cite{Armbruster:2002mp} and 
the NOMAD Collaboration~\cite{Astier:2003gs} are also shown. We note that the excursion to small mixing 
in the exclusion contour at around $\Delta m^2_{41} \sim 5\times 10^{-3}$\,\,eV$^2$ is 
originated from the island in Fig.~\ref{fig:minoscls}.}
\end{figure}

In our combined analysis, we obtain $\Delta \chi^2_\mathrm{obs}$ as well as $\Delta \chi^2_{3\nu}$ and 
$\Delta \chi^2_{4\nu}$ distributions for each $(\sin^2 2\theta_{14}, 
\Delta m^2_{41})$ grid point of the Daya Bay and Bugey-3 combination, 
and for each $(\sin^2 \theta_{24}, \Delta m^2_{41})$ grid point from the MINOS experiment.
We then combine pairs of grid points from the MINOS and the Daya Bay and Bugey-3 results at fixed values of $\Delta m^2_{41}$ to obtain constraints on electron neutrino or antineutrino appearance due to oscillations into sterile neutrinos.
Since the systematic uncertainties of accelerator and reactor experiments are largely uncorrelated, for each $(\sin^2 2\theta_{14}, \sin^2\theta_{24}, \Delta m^2_{41})$ grid point,  a combined $\Delta \chi^2_\mathrm{obs}$ is constructed from the sum of the corresponding MINOS and Daya Bay/Bugey-3 $\Delta \chi^2_\mathrm{obs}$ values. 
Similarly, the combined $\Delta \chi^2_{3\nu}$ and $\Delta \chi^2_{4\nu}$ 
distributions are constructed by adding random samples drawn from the corresponding MINOS and Daya Bay/Bugey-3 
distributions. Finally, the  \cls\ value at every $(\sin^2 2\theta_{14},\sin^2 \theta_{24})$ point is 
calculated using Eq.~\eqref{eq:clsDef}, while the $\Delta m^2_{41}$ value is fixed. 
While \cls\ is single valued  at every $(\sin^2 2\theta_{14},\sin^2 \theta_{24})$ point for a given value of $\Delta m^2_{41}$, it is  multivalued as a function of $\sin^2 2\theta_{\mu e}$ ({\it cf.} Eq.~\eqref{eq:DisapToApp}). 
To obtain a single-valued function, we make the conservative choice of selecting the largest \cls\ value for any given $\sin^2 2\theta_{\mu e}$.  
The 90\% CL$_s$ exclusion contour resulting from this procedure is shown in Fig.~\ref{fig:combocls}. Under the assumption of {\it CPT} conservation, the combined constraints are equally valid in constraining electron neutrino or antineutrino appearance.
The combined results of the Daya Bay + Bugey-3 and the MINOS experiments constrain $\sin^2 2\theta_{\mu e} <$~[\num{3.0e-4}~(90\% CL$_s$), \num{4.5e-4}~(95\% CL$_s$)] for $\Delta m^2_{41}$~$=$~1.2~eV$^2$. 

In conclusion, we have combined constraints on $\sin^2 2\theta_{14}$ derived from a search for electron antineutrino disappearance at the Daya Bay and Bugey-3 reactor experiments with constraints on $\sin^2 \theta_{24}$ derived from a search for muon (anti)neutrino disappearance in the NuMI beam at the MINOS experiment. Assuming a four-flavor model of active-sterile oscillations, we constrain $\sin^2 2\theta_{\mu e}$, the parameter controlling electron (anti)neutrino appearance at short-baseline experiments, over 6 orders of magnitude in $\Delta m^2_{41}$. We set the strongest constraint to date and exclude the sterile neutrino mixing phase space allowed by the LSND and MiniBooNE experiments for $\Delta m^2_{41} < 0.8$~eV$^2$ at a 95\% \cls. Our results are in good agreement with results from global fits (see Refs.~\cite{Kopp:2013vaa,Gariazzo:2015rra} and references therein) at specific parameter choices; however, they differ in detail over the range of parameter space.  The results explicitly show the strong tension between null results from disappearance searches and appearance-based indications for the 
existence of light sterile neutrinos.

The MINOS experiment is supported by the U.S. Department of Energy, the United Kingdom Science and Technology Facilities Council, the U.S. National Science Foundation, the State and University of Minnesota, and Brazil's FAPESP (Funda\c{c}\~{a}o de Amparo \`{a} Pesquisa do Estado de S\~{a}o Paulo), CNPq (Conselho Nacional de Desenvolvimento Cient\'{i}fico e Tecnol\'{o}gico), and CAPES (Coordena\c{c}\~{a}o de Aperfei\c{c}oamento de Pessoal de N\'{i}vel Superior).  We are grateful to the Minnesota Department of Natural Resources and the personnel of the Soudan Laboratory and Fermilab. We thank the Texas Advanced Computing Center at The University of Texas at Austin for the provision of computing resources.

The Daya Bay experiment is supported in part by 
the Ministry of Science and Technology of China,
the U.S. Department of Energy,
the Chinese Academy of Sciences,
the CAS Center for Excellence in Particle Physics,
the National Natural Science Foundation of China,
the Guangdong provincial government,
the Shenzhen municipal government,
the China General Nuclear Power Group,
the Research Grants Council of the Hong Kong Special Administrative Region of China,
the Ministry of Education in Taiwan,
the U.S. National Science Foundation,
the Ministry of Education, Youth and Sports of the Czech Republic,
the Joint Institute of Nuclear Research in Dubna, Russia,
the NSFC-RFBR joint research program,
and the National Commission for Scientific and Technological Research of Chile.
We acknowledge Yellow River Engineering Consulting Co., Ltd.\ and China Railway 15th Bureau Group Co., Ltd.\ for building the underground laboratory.
We are grateful for the ongoing cooperation from the China Guangdong Nuclear Power Group and China Light~\&~Power Company.

\vskip 0.1 true cm
{\it Note Added}.\---Recently, a paper appeared by the IceCube Collaboration that sets limits using sterile-driven disappearance of muon neutrinos~\cite{TheIceCube:2016oqi}. The results place strong constraints on $\sin^2 2\theta_{24}$ for $\Delta m^2_{41} \in (0.1,10)\ {\rm eV}^2$. Further, a paper that reanalyses the same IceCube data in a model including nonstandard neutrino interactions also appeared~\cite{Liao:2016reh}.

\bibliographystyle{apsrev4-1}
\bibliography{dyb-minos-sterile}

\end{document}

%% file: merged_minos_dyb_authors.tex
%
\newcommand{\NUU}{{National~United~University, Miao-Li}}
\newcommand{\Minnesota}{University of Minnesota, Minneapolis, Minnesota 55455, USA}
\newcommand{\ECUST}{{Institute of Modern Physics, East China University of Science and Technology, Shanghai}}
\newcommand{\SJTU}{{Department of Physics and Astronomy, Shanghai Jiao Tong University, Shanghai Laboratory for Particle Physics and Cosmology, Shanghai}}
\newcommand{\Charles}{{Charles~University, Faculty~of~Mathematics~and~Physics, Prague, Czech~Republic}} 
\newcommand{\JMU}{Physics Department, James Madison University, Harrisonburg, Virginia 22807, USA}
\newcommand{\Delhi}{Department of Physics \& Astrophysics, University of Delhi, Delhi 110007, India}
\newcommand{\NJU}{{Nanjing~University, Nanjing}}
\newcommand{\LosAlamos}{Los Alamos National Laboratory, Los Alamos, New Mexico 87545, USA}
\newcommand{\Rochester}{Department of Physics and Astronomy, University of Rochester, New York 14627 USA}
\newcommand{\Iowa}{Department of Physics and Astronomy, Iowa State University, Ames, Iowa 50011 USA}
\newcommand{\Athens}{Department of Physics, University of Athens, GR-15771 Athens, Greece}
\newcommand{\XJTU}{{Xi'an Jiaotong University, Xi'an}}
\newcommand{\SLAC}{Stanford Linear Accelerator Center, Stanford, California 94309, USA}
\newcommand{\Yale}{{Department~of~Physics, Yale~University, New~Haven, Connecticut 06520, USA}}
\newcommand{\NTU}{{Department of Physics, National~Taiwan~University, Taipei}}
\newcommand{\PennU}{Department of Physics and Astronomy, University of Pennsylvania, Philadelphia, Pennsylvania 19104, USA}
\newcommand{\CdF}{APC -- Universit\'{e} Paris 7 Denis Diderot, 10, rue Alice Domon et L\'{e}onie Duquet, F-75205 Paris Cedex 13, France}
\newcommand{\Oxford}{Subdepartment of Particle Physics, University of Oxford, Oxford OX1 3RH, United Kingdom}
\newcommand{\Caltech}{Lauritsen Laboratory, California Institute of Technology, Pasadena, California 91125, USA}
\newcommand{\Alabama}{Department of Physics and Astronomy, University of Alabama, Tuscaloosa, Alabama 35487, USA}
\newcommand{\Houston}{Department of Physics, University of Houston, Houston, Texas 77204, USA}
\newcommand{\SZU}{{Shenzhen~University, Shenzhen}}
\newcommand{\LLL}{Lawrence Livermore National Laboratory, Livermore, California 94550, USA}
\newcommand{\PennState}{Department of Physics, Pennsylvania State University, State College, Pennsylvania 16802, USA}
\newcommand{\Washington}{Physics Department, Western Washington University, Bellingham, Washington 98225, USA}
\newcommand{\HKU}{{Department of Physics, The~University~of~Hong~Kong, Pokfulam, Hong~Kong}}
\newcommand{\TsingHua}{{Department~of~Engineering~Physics, Tsinghua~University, Beijing}}
\newcommand{\LASL}{Nuclear Nonproliferation Division, Threat Reduction Directorate, Los Alamos National Laboratory, Los Alamos, New Mexico 87545, USA}
\newcommand{\Indiana}{Indiana University, Bloomington, Indiana 47405, USA}
\newcommand{\deceased}{Deceased.}
\newcommand{\Otterbein}{Otterbein University, Westerville, Ohio 43081, USA}
\newcommand{\UIUC}{{Department of Physics, University~of~Illinois~at~Urbana-Champaign, Urbana, Illinois 61801, USA}}
\newcommand{\TempleUniversity}{{Department~of~Physics, College~of~Science~and~Technology, Temple~University, Philadelphia, Pennsylvania  19122, USA}}
\newcommand{\RAL}{Rutherford Appleton Laboratory, Science and Technology Facilities Council, Didcot, OX11 0QX, United Kingdom}
\newcommand{\ZSU}{{Sun Yat-Sen (Zhongshan) University, Guangzhou}}
\newcommand{\BCC}{{Now at Department of Chemistry and Chemical Technology, Bronx Community College, Bronx, New York  10453, USA}} 
\newcommand{\Manchester}{School of Physics and Astronomy, University of Manchester, Manchester M13 9PL, United Kingdom}
\newcommand{\RPI}{{Department~of~Physics, Applied~Physics, and~Astronomy, Rensselaer~Polytechnic~Institute, Troy, New~York  12180, USA}}
\newcommand{\GEHealth}{GE Healthcare, Florence South Carolina 29501, USA}
\newcommand{\SDU}{{Shandong~University, Jinan}}
\newcommand{\Crookston}{Math, Science and Technology Department, University of Minnesota -- Crookston, Crookston, Minnesota 56716, USA}
\newcommand{\Tufts}{Physics Department, Tufts University, Medford, Massachusetts 02155, USA}
\newcommand{\CQU}{{Chongqing University, Chongqing}} 
\newcommand{\Cleveland}{Cleveland Clinic, Cleveland, Ohio 44195, USA}
\newcommand{\ITEP}{High Energy Experimental Physics Department, ITEP, B. Cheremushkinskaya, 25, 117218 Moscow, Russia}
\newcommand{\TechX}{Tech-X Corporation, Boulder, Colorado 80303, USA}
\newcommand{\Harvard}{Department of Physics, Harvard University, Cambridge, Massachusetts 02138, USA}
\newcommand{\Sussex}{Department of Physics and Astronomy, University of Sussex, Falmer, Brighton BN1 9QH, United Kingdom}
\newcommand{\Benedictine}{Physics Department, Benedictine University, Lisle, Illinois 60532, USA}
\newcommand{\Lebedev}{Nuclear Physics Department, Lebedev Physical Institute, Leninsky Prospect 53, 119991 Moscow, Russia}
\newcommand{\Wisconsin}{Physics Department, University of Wisconsin, Madison, Wisconsin 53706, USA}
\newcommand{\Cambridge}{Cavendish Laboratory, University of Cambridge,Cambridge CB3 0HE, United Kingdom}
\newcommand{\BNL}{Brookhaven National Laboratory, Upton, New York 11973, USA}
\newcommand{\NUDT}{{College of Electronic Science and Engineering, National University of Defense Technology, Changsha}} 
\newcommand{\BNU}{{Beijing~Normal~University, Beijing}}
\newcommand{\NanKai}{{School of Physics, Nankai~University, Tianjin}}
\newcommand{\SDakota}{South Dakota School of Mines and Technology, Rapid City, South Dakota 57701, USA}
\newcommand{\DGUT}{{Dongguan~University~of~Technology, Dongguan}}
\newcommand{\UC}{{Department of Physics, University~of~Cincinnati, Cincinnati, Ohio 45221, USA}}
\newcommand{\UFG}{Instituto de F\'{i}sica, Universidade Federal de Goi\'{a}s, 74690-900, Goi\^{a}nia, GO, Brazil}
\newcommand{\IHEP}{Institute for High Energy Physics, Protvino, Moscow Region RU-140284, Russia}
\newcommand{\Princeton}{{Joseph Henry Laboratories, Princeton~University, Princeton, New~Jersey 08544, USA}}
\newcommand{\UNICAMP}{Universidade Estadual de Campinas, IFGW, CP 6165, 13083-970, Campinas, SP, Brazil}
\newcommand{\Carolina}{Department of Physics and Astronomy, University of South Carolina, Columbia, South Carolina 29208, USA}
\newcommand{\Siena}{{Siena~College, Loudonville, New York  12211, USA}}
\newcommand{\RoyalH}{Physics Department, Royal Holloway, University of London, Egham, Surrey, TW20 0EX, United Kingdom}
\newcommand{\Warsaw}{Department of Physics, University of Warsaw,PL-02-093 Warsaw, Poland}
\newcommand{\UCB}{{Department of Physics, University~of~California, Berkeley, California  94720, USA}}
\newcommand{\UCL}{Department of Physics and Astronomy, University College London, London WC1E 6BT, United Kingdom}
\newcommand{\IIT}{Department of Physics, Illinois Institute of Technology, Chicago, Illinois 60616, USA}
\newcommand{\VirginiaTech}{{Center for Neutrino Physics, Virginia~Tech, Blacksburg, Virginia  24061, USA}}
\newcommand{\Texas}{Department of Physics, University of Texas at Austin, Austin, Texas 78712, USA}
\newcommand{\USP}{Instituto de F\'{i}sica, Universidade de S\~{a}o Paulo,  CP 66318, 05315-970, S\~{a}o Paulo, SP, Brazil}
\newcommand{\Pittsburgh}{Department of Physics and Astronomy, University of Pittsburgh, Pittsburgh, Pennsylvania 15260, USA}
\newcommand{\NTUAthens}{Department of Physics, National Tech. University of Athens, GR-15780 Athens, Greece}
\newcommand{\USTC}{{University~of~Science~and~Technology~of~China, Hefei}}
\newcommand{\CIAE}{{China~Institute~of~Atomic~Energy, Beijing}}
\newcommand{\Cincinnati}{Department of Physics, University of Cincinnati, Cincinnati, Ohio 45221, USA}
\newcommand{\FNAL}{Fermi National Accelerator Laboratory, Batavia, Illinois 60510, USA}
\newcommand{\NCTU}{{Institute~of~Physics, National~Chiao-Tung~University, Hsinchu}}
\newcommand{\Duluth}{Department of Physics, University of Minnesota Duluth, Duluth, Minnesota 55812, USA}
\newcommand{\TexasAM}{Physics Department, Texas A\&M University, College Station, Texas 77843, USA}
\newcommand{\Berkeley}{Lawrence Berkeley National Laboratory, Berkeley, California, 94720 USA}
\newcommand{\HolyCross}{Holy Cross College, Notre Dame, Indiana 46556, USA}
\newcommand{\MIT}{Lincoln Laboratory, Massachusetts Institute of Technology, Lexington, Massachusetts 02420, USA}
\newcommand{\CGNPG}{{China General Nuclear Power Group}}
\newcommand{\ANL}{Argonne National Laboratory, Argonne, Illinois 60439, USA}
\newcommand{\Dubna}{{Joint~Institute~for~Nuclear~Research, Dubna, Moscow~Region}}
\newcommand{\CUHK}{{Chinese~University~of~Hong~Kong, Hong~Kong}}
\newcommand{\NCEPU}{{North~China~Electric~Power~University, Beijing}}
\newcommand{\CNIHEP}{{Institute~of~High~Energy~Physics, Beijing}}
\newcommand{\Stanford}{Department of Physics, Stanford University, Stanford, California 94305, USA}
\newcommand{\Lancaster}{Lancaster University, Lancaster, LA1 4YB, UK}
\newcommand{\Ohio}{Center for Cosmology and Astro Particle Physics, Ohio State University, Columbus, Ohio 43210 USA}
\newcommand{\StJohnFisher}{Physics Department, St. John Fisher College, Rochester, New York 14618 USA}
\newcommand{\CUC}{{Instituto de F\'isica, Pontificia Universidad Cat\'olica de Chile, Santiago, Chile}} 
\newcommand{\WandM}{Department of Physics, College of William \& Mary, Williamsburg, Virginia 23187, USA}
\author{P.~Adamson\ensuremath{^{\mu}}}
\affiliation{\FNAL} 
\author{F.~P.~An\ensuremath{^{\delta}}}
\affiliation{\ECUST} 
\author{I.~Anghel\ensuremath{^{\mu}}}
\affiliation{\Iowa}\affiliation{\ANL} 
\author{A.~Aurisano\ensuremath{^{\mu}}}
\affiliation{\Cincinnati} 
\author{A.~B.~Balantekin\ensuremath{^{\delta}}}
\affiliation{\Wisconsin} 
\author{H.~R.~Band\ensuremath{^{\delta}}}
\affiliation{\Yale} 
\author{G.~Barr\ensuremath{^{\mu}}}
\affiliation{\Oxford} 
\author{M.~Bishai\ensuremath{^{\delta}}\ensuremath{^{\mu}}}
\affiliation{\BNL} 
\author{A.~Blake\ensuremath{^{\mu}}}
\affiliation{\Cambridge}\affiliation{\Lancaster} 
\author{S.~Blyth\ensuremath{^{\delta}}}
\affiliation{\NTU}\affiliation{\NUU} 
% ...
\collaboration{\ensuremath{^{\delta}}Daya Bay Collaboration}\noaffiliation
\collaboration{\ensuremath{^{\mu}}MINOS Collaboration}\noaffiliation